# SUPERPARAMAGNETISM IN NANOCRYSTALLINE COPPER FERRITE THIN FILMS

Prasanna D. Kulkarni[1], R.P.R.C. Aiyar[2], Shiva Prasad[1], N. Venkataramani[2], R. Krishnan[3], Wenjie Pang[4], Ayon Guha[4], R.C. Woodward[4], R.L. Stamps[4].
(1) Physics Department, IIT Bombay, Powai, Mumbai 400076, (India) (2) ACRE, IIT Bombay, Powai, Mumbai 400076 (India), (3) Laboratoire de Magnetisme et d'optique de Versailles, CNRS, 78935 Versailles (France), (4) School of Physics, M013, The University of Western Australia, 35 Stirling Hwy, Crawley WA 6009 (Australia).

Keywords: Ferrimagnetic thin films, Nanocrystalline, Ferrimagnetism, Superparamagnetism

## ABSTRACT

The rf sputtered copper ferrite films contain nanocrystalline grains. In these films, the magnetization does not saturate even in high magnetic fields. This phenomenon of high field susceptibility is attributed to the defects and/or superparamagnetic grains in the films. A simple model is developed to describe the observed high field magnetization behavior of these films. The model is found to fit well to the high field part for all the studied films. An attempt is also made to explain the temperature variation of the ferrimagnetic contribution on the basis of reported exchange constants.

## INTRODUCTION

One of the observations made in a number of ferrite thin film studies is the departure of the magnetization from the established bulk value and non saturation even under high magnetic fields. The deviation from the known bulk value of magnetization is attributed to the presence of nanocrystallinity and change in cation distribution. The non saturation aspect is postulated to arise from local defects, anisotropy and superparamagnetic grains. This is discussed earlier [1-3] on the basis of approach to saturation of magnetization using the Chikazumi expression [4], where the high field susceptibility (HFS) is related to the defects present in the films. However, it is not possible to completely understand the temperature dependence of the high field susceptibility, using this model in the case of copper ferrite thin films. Because for room temperature as deposited films, the high field susceptibility is found to increase when the temperature increases from 5K to 300K. If the HFS is purely related to the anisotropic effects, it is expected to decrease with increasing temperature. The rf sputtered copper ferrite thin films have nanocrystalline grains, therefore there is a likelihood of a large fraction of particles falling in the superparamagnetic regime. In such a case, the magnetic moments of the nanocrystalline particles are unable to align in the field direction because of the dominant thermal fluctuations. Bean and Livingston [5] have discussed a critical volume for particles to become superparamagnetic in zero field. Above this critical volume, the particles are stable at a given temperature.

In this paper, we report the superparamagnetism in rf sputtered copper ferrite films using a theoretical model. A similar type of work on amorphous ribbons is reported by Wang et. al. [6].

Copper ferrite can be stabilized in two different phases in thin film form, viz., a cubic and a tetragonal phase [7]. Quenching the copper ferrite films after carrying out a post deposition annealing stabilizes the cubic phase. The slow cooling of the films after the annealing, on the other hand, results in the tetragonal phase. High field magnetization studies are carried out for as deposited and annealed films of copper ferrite.

**EXPERIMENTAL**

Copper ferrite films are deposited using a Leybold Z400 rf sputtering system on amorphous quartz substrates. No heating or cooling is carried out during the sputter deposition of the films. The rf power employed during the deposition of the films reported in the present study is 200W. The thickness of the films is ~2400 Å. The films are subsequently annealed in air at 800°C for 2 hours, followed by either slow cooling or quenching in liquid nitrogen. The temperature dependence of magnetic properties is studied using a SQUID magnetometer for as deposited (Asd), slow cooled (SC) and quenched (Q) copper ferrite thin films. M vs T data is recorded in temperature range 5K to 300K at a field of 3T for Asd and SC film and at 1T for the Q film. The M vs H curves are traced at various temperatures from 5K to 300K at a field upto 7T.

**RESULTS**

The normalized M vs. T curves for the Asd, SC and Q films are shown in Fig.1. The substrate contribution is subtracted from the total magnetization. The data is normalized with respect to the corresponding magnetization at 5K. The M vs T curve shows a faster decrease for Asd and Q films than for the SC film. This shows that temperature dependence of magnetization is different for the two phases of copper ferrite. The M vs. T data is fitted to a theoretical M vs. T curve of copper ferrite, based on the calculations of Srivastava et. al. [8]. These calculations are an extension of Neel's model for cubic inverse spinel ferrite. It is observed that the M vs. T curves of the films are not described adequately by the fittings. The calculated M values are lower than the M values observed for the copper ferrite thin films. The reason could be the field at which M vs. T curves are recorded does not saturate the magnetization in copper ferrite thin films. Therefore, there is a contribution to the total magnetization from the non saturating part of the M vs H curve in the case of nanocrystalline systems.

The rf sputtered copper ferrite thin films are nanocrystalline with a wide grain size distribution (Asd - 5nm–25nm, SC and Q - 10nm-100nm) [7]. A fraction of these grains may fall in the superparamagnetic regime. The superparamagnetic grains may contribute significantly to the net magnetization of the films.

We have developed a simple model to take into account the contribution of the high field part of M vs. H curve to the total magnetization of the film. In this model, as a first approximation, the superparamagnetic grains at each temperature are replaced by non interacting clusters. These clusters, which represent the superparamagnetic grains, show substantially large moment in the applied field. The average moment with each cluster is assumed to be the same. The magnetization of these clusters is then expected to follow a Langevin function [5]. However, the total magnetization which is experimentally observed, is the sum of the contributions from both superparamagnetic and ferrimagnetic particles. This may be represented as,

$$M_T = M_{FE} + M_{SP} \tag{1}$$

where $M_{SP}$ represents the total magnetic moment due to superparamagnetic contribution and $M_{FE}$ represents the ferrimagnetic contribution to the total magnetization.

The high field part of the M Vs H curve can be fitted to the expression,

$$M = M_{FE}\,[\,1 + a\,L(\mu H/kT)\,] \tag{2}$$

Here, $M_{FE}$ is the saturation magnetization of the ferrimagnetic part of the M vs H curve, μ is the average magnetic moment per superparamagnetic cluster in terms of Bohr magneton, $k$ is the Boltzmann constant, $T$ is the temperature. The parameter '*a*' is the ratio of superparamagnetic contribution to the ferrimagnetic contribution of the magnetization. The Eqn (2) is found to fit to the high field part of magnetization for Asd, SC and Q films at all temperatures. The results of these fittings for Asd, SC and Q films are shown in Fig. 2, only for the 5K and 300K data. From these fittings the parameters μ, '*a*' and $M_{FE}$ are determined. Fig.3 shows the temperature dependence of μ for Asd, SC and Q films. The temperature dependence of normalized magnetization for the ferrimagnetic part for the Q film is shown in Fig 4, along with the normalized theoretical M vs. T curve.

## DISCUSSION

The average magnetic moment per superparamagnetic cluster reduces with temperature as shown in Fig. 3. This is due to the freezing in of the larger superparamagnetic clusters when the temperature is reduced. These 'blocked' clusters then contribute to the ferrimagnetic part of the magnetization. The smaller sized clusters continue to contribute to the superparamagnetic part of the magnetization at lower temperature. The numerical value of μ for Asd film is lower than the corresponding μ value for annealed films. This shows a presence of smaller grain sizes in the Asd film. The grain size increases when the films are ex-situ annealed and this is reflected in the higher value of μ for the annealed films.

The temperature dependence of ferrimagnetic part shows that $M_{FE}$ increases with decrease in temperature for the cubic phase Q film, as seen in Fig. 4. For the cubic phase Q film having the largest magnetization, the $M_{FE}$ vs T follows the theoretical M vs. T curve obtained from Srivastava et. al. [8]. However, for Asd and SC films, the temperature dependence of magnetization is not properly described by the ferrimagnetic component of the magnetization obtained by Eq. (2). The parameter '*a*' is also obtained from the fittings of Eq. (2) to the high field magnetization data. This parameter represents the ratio of superparamagnetic contribution and the ferromagnetic contribution to the total magnetization. Because of blocking of the superparamagnetic particles at lower temperatures, the value of '*a*' is expected to decrease as the temperature is reduced for all the three films of copper ferrite. However, the variation of '*a*' is also observed to be inconsistent as the temperature is lowered. The simple model, presented in this paper, work well for obtaining the cluster moments. But, it needs modification to determine the ferrimagnetic contributions to the total magnetization in the case of Asd and SC film by considering the particles to be interacting and modifying the Langevin function accordingly [6]. A refinement of the simple model is under progress.

## CONCLUSION

A model involving the Langevin function describes the high field magnetization behavior of nanocrystalline copper ferrite thin films at all temperatures. The magnetic moment of the superparamagnetic clusters reduces with temperature due to freezing of the moment at lower temperatures. For the cubic phase Q film, the model works well in separating the ferrimagnetic and superparamagnetic contributions to the total magnetization of the film.

**LIST OF FIGURES:**

**Fig.1.** Temperature dependence of normalized magnetization of SC, Q and Asd copper ferrite films.

**Fig.2.** High field part of the magnetization fitted with Eqn (2) for (a) SC, (b) Q, (c) Asd copper ferrite films.

**Fig.3.** Temperature dependence of superparamagnetic cluster moment for SC, Q and Asd copper ferrite films.

**Fig.4.** Temperature dependence of the ferrimagnetic component of the magnetization for the Q film.

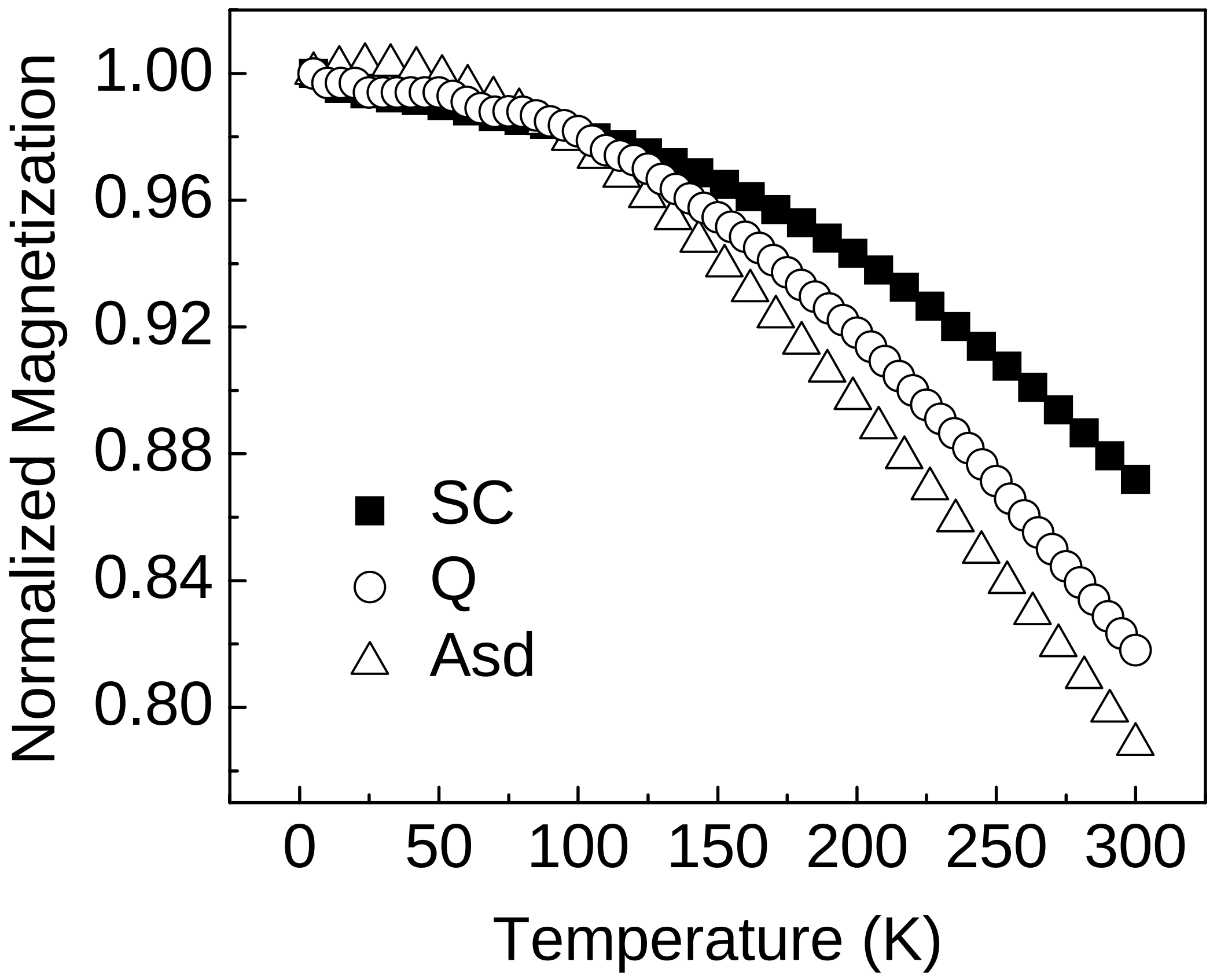

Fig.1.

Prasanna D. Kulkarni et. al. “SUPERPARAMAGNETISM IN NANOCRYSTALLINE COPPER FERRITE THIN FILMS”

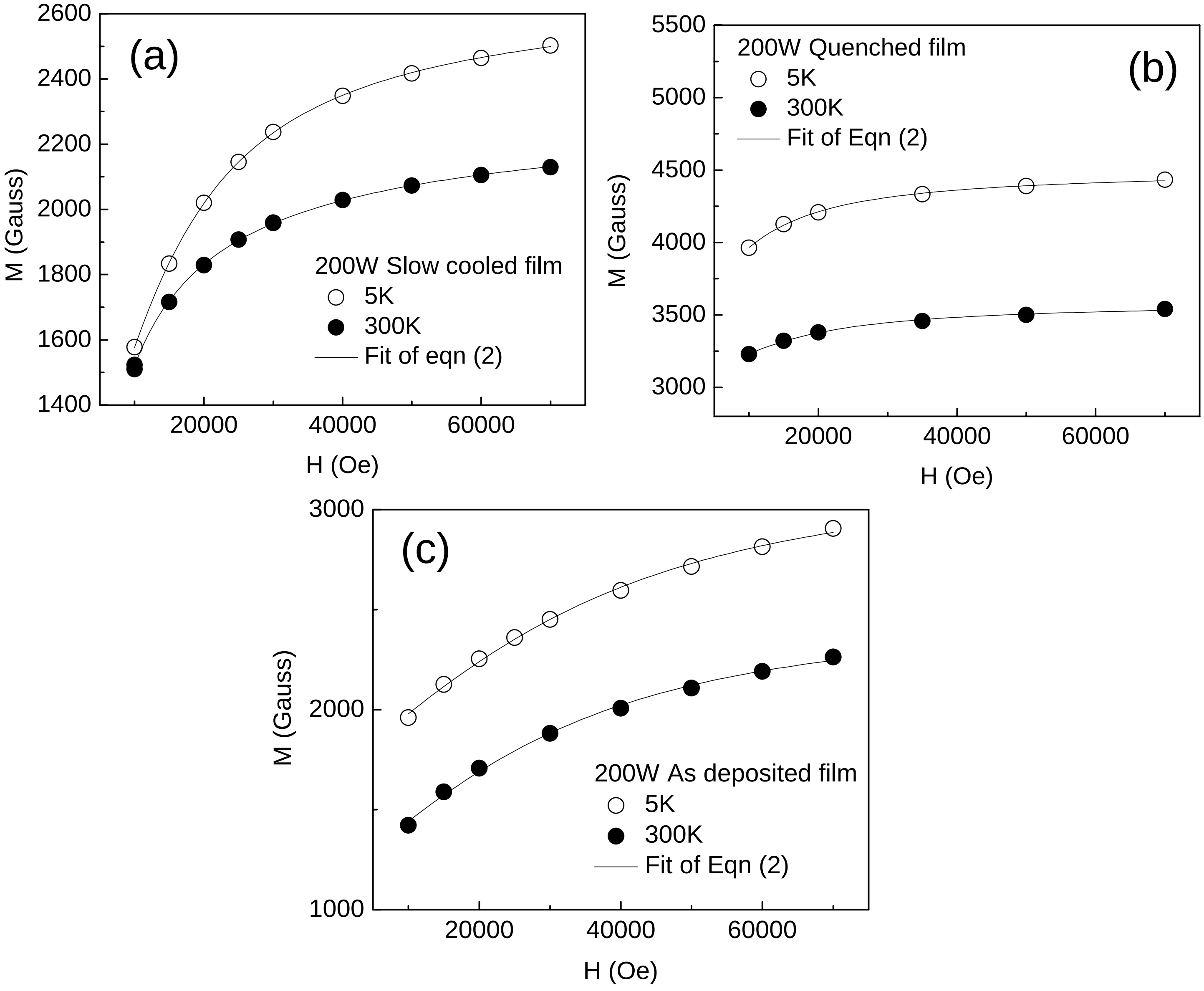

Fig.2.

Prasanna D. Kulkarni et. al. "SUPERPARAMAGNETISM IN NANOCRYSTALLINE COPPER FERRITE THIN FILMS"

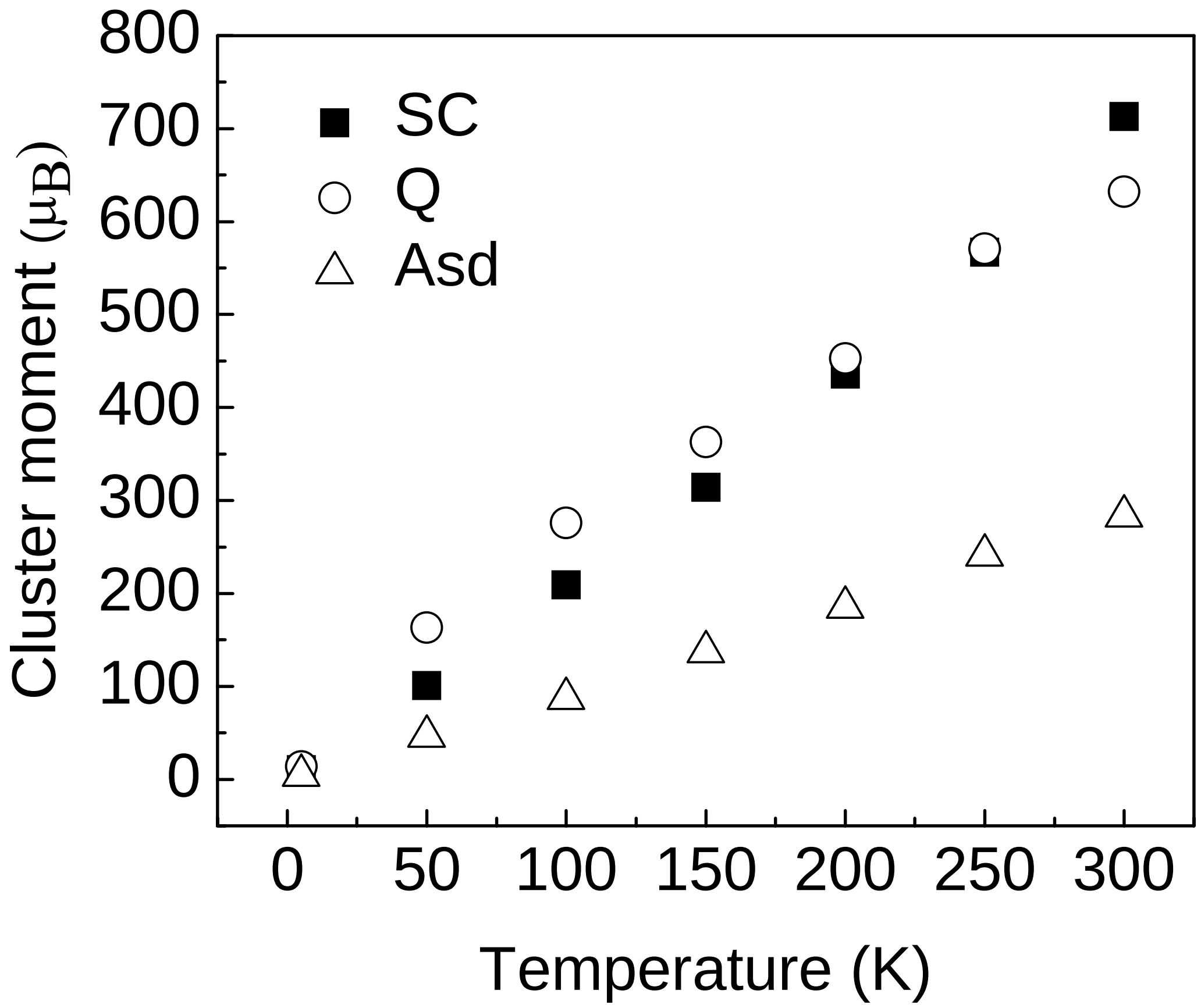

Fig.3.

Prasanna D. Kulkarni et. al. “SUPERPARAMAGNETISM IN NANOCRYSTALLINE COPPER FERRITE THIN FILMS”

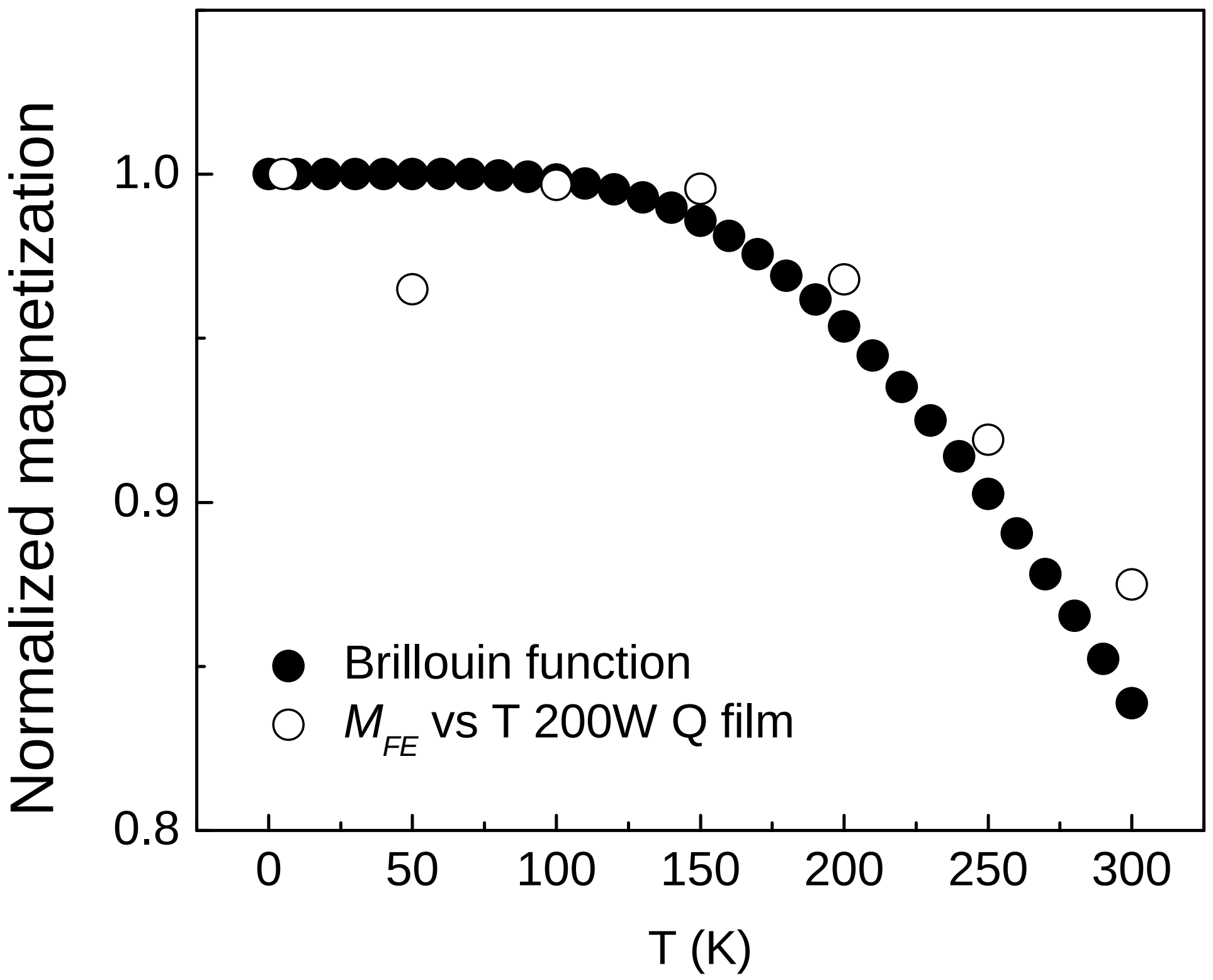

Fig.4.

Prasanna D. Kulkarni et. al. “SUPERPARAMAGNETISM IN NANOCRYSTALLINE COPPER FERRITE THIN FILMS”